\documentclass[10pt]{article}

\usepackage{amsmath}
\usepackage{amssymb}
\usepackage{graphicx}
\usepackage{cite}
\usepackage{color} 
 
\usepackage[labelfont=bf,labelsep=period,justification=raggedright]{caption}

\makeatletter
\renewcommand{\@biblabel}[1]{\quad#1.}
\makeatother

\date{}

\begin{document}

\begin{flushleft}
{\Large
\textbf{Geographic constraints on social network groups}
}
\\
Jukka-Pekka Onnela$^{1\dagger,\ast}$, 
Samuel Arbesman$^{1\dagger}$, 
Marta C. Gonz\'{a}lez$^{2}$,
 Albert-L\'{a}szl\'{o} Barab\'{a}si$^{3,4,5}$
Nicholas A. Christakis$^{1,6,7}$
\\
\bf{1} Department of Health Care Policy, Harvard Medical School, Boston, MA, USA
\\
\bf{2} Department of Civil and Environmental Engineering and Engineering Systems, Massachusetts Institute of Technology, Cambridge, MA, USA
\\
\bf{3} Center for Complex Network Research, Department of Physics, Biology and Computer Science, Northeastern University, Boston, MA, USA
\\
\bf{4} Center for Cancer Systems Biology, Dana Farber Cancer Institute, Boston, MA, USA
\\
\bf{5} Department of Medicine, Brigham and WomenÕs Hospital, Harvard Medical School, MA, USA
\\
\bf{6} Department of Medicine, Harvard Medical School, Boston, MA, USA
\\
\bf{7} Department of Sociology, Harvard Faculty of Arts and Sciences, Cambridge, MA, USA
\\
$\ast$ E-mail: onnela@med.harvard.edu
\\
$\dagger$ These authors contributed equally to this work.
\end{flushleft}

\section*{Abstract}
Social groups are fundamental building blocks of human societies. While our social interactions have always been constrained by geography, it has been impossible, due to practical difficulties, to evaluate the nature of this restriction on social group structure. We construct a social network of individuals whose most frequent geographical locations are also known. We also classify the individuals into groups according to a community detection algorithm. We study the variation of geographical span for social groups of varying sizes, and explore the relationship between topological positions and geographic positions of their members. We find that small social groups are geographically very tight, but become much more clumped when the group size exceeds about 30 members. Also, we find no correlation between the topological positions and geographic positions of individuals within network communities. These results suggest that spreading processes face distinct structural and spatial constraints.

\section*{Introduction}
Social groups are common among animals and humans\cite{newman,caldarelli,mendes,albert,structure}. In humans, they reflect friendship, kinship, and work relationships, and can also be seen as social networks. From an evolutionary and historical perspective, the formation of such network groups -- consisting of agglomerations of dyadic interactions -- has been constrained by geography. In contrast, larger social units, enabled by modern technology and political organization, offer drastically different opportunities for social interactions and for group assembly over larger geographic ranges. This raises two sorts of questions. First, is the structure of ``old-fashioned" groups similar to the large-scale groups possible in modern society? And second, what role does geography play in group formation?

If we represent the social relationships among a population of people as a network, then groups can be seen as ``communities'' within the population that consist of sets of nodes that are relatively densely connected to each other but sparsely connected to other nodes in the network\cite{commreview,fortunato}. While social communities have been studied for a long time\cite{freemanbook}, it has recently become feasible, with mobile phone data, to monitor the social interactions and geographic positions of millions of individuals\cite{onnela,gonzalez}, and to apply algorithmic detection of communities on a large scale\cite{commreview,fortunato}. The structure of dyadic social interactions is known to depend on geography, for example, as shown by the decay of friendship probability with distance, based on voluntary self-reports of hometown and US state, in a blog community\cite{libennowell}, and the decrease in communication probability with distance based on the zip codes of cell phone billing addresses\cite{lambiotte}. In addition, a previous study has shown that smaller communities are more homogeneous with respect to the billing postal codes of their members\cite{palla}, while another presented evidence that this persists across a hierarchy of communities \cite{ahn}. However, there are no prior large-scale studies of the way in which community structure depends on geography, where the actual communication locations are used and where geographical properties of communities themselves are examined (see Fig.~1).

\begin{figure}[!ht]
\begin{center}
\includegraphics[width=0.55\linewidth]{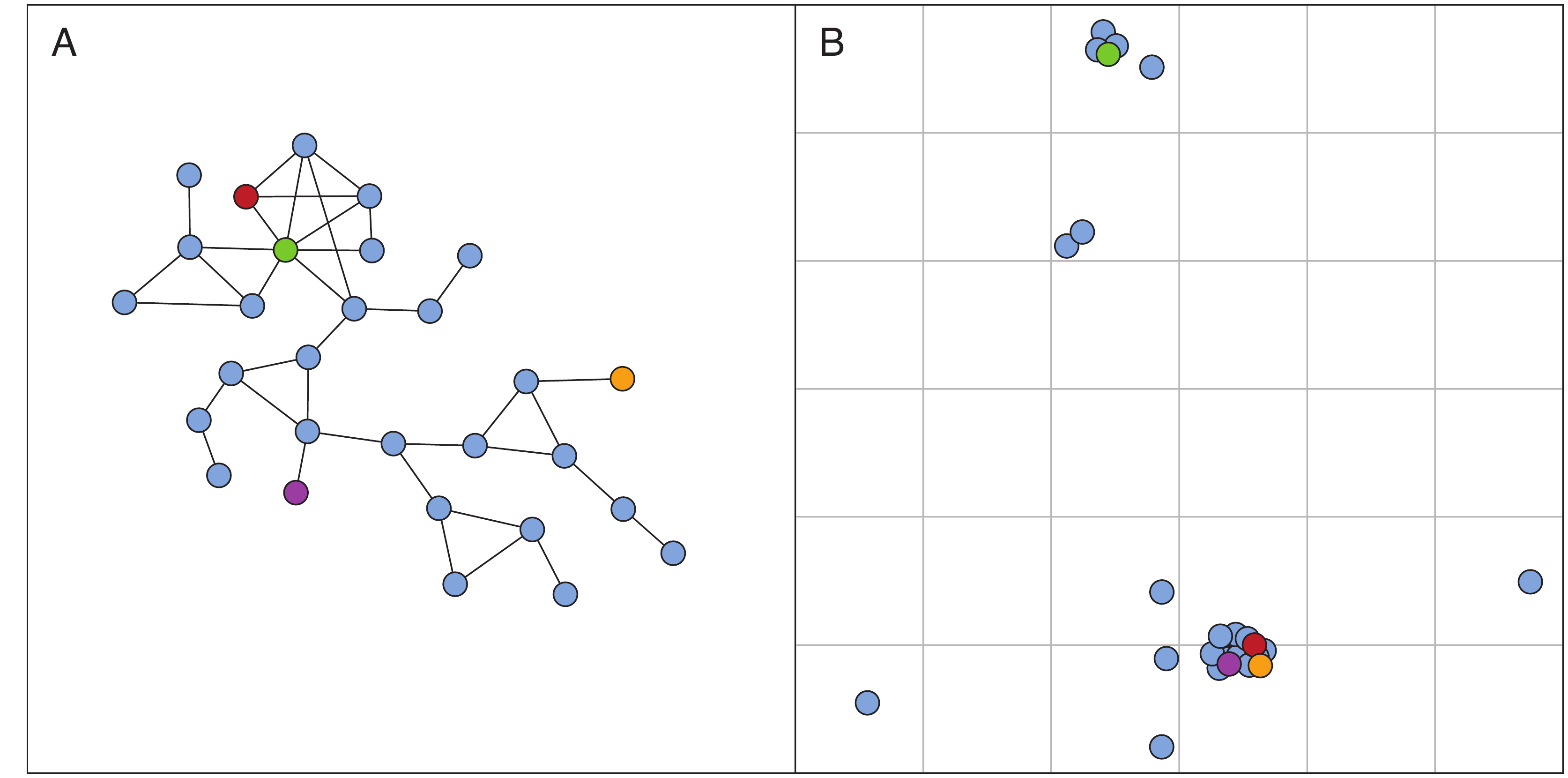}
\end{center}
\caption{{\bf Visualization of a community in the mobile phone network.} This juxtaposition of (\textbf{A}) the topological structure and (\textbf{B}) the geographical structure demonstrates the interplay of these two dimensions. The purple and orange nodes are geographically close, but topologically they lie at five degrees of separation. In contrast, the red and green nodes are connected to each other, and also share several neighbors, yet they are geographically separated by a large distance. Overlapping nodes in (B) have been moved slightly for visual clarity.}
\end{figure}

With respect to group formation, geography can be seen as a kind of constraint. That is, social connections not only face network constraints and opportunities (we tend to form ties with others who are the friends of our friends), but also, quite obviously, geographic constraints and opportunities.  What is unclear, however, is the way in which such geographic constraints and opportunities affect and shape network communities above and beyond their effect on dyadic interactions.

\section*{Results}
\subsection*{Dyadic Interactions and Geography}

We create a network of social interactions by measuring ties between individuals based on mobile phone call and text messaging data from an unnamed European country. Based on the records of 72.4 millions calls and 17.1 million text messages accumulated over a one-month period, the resulting network has 3.4 million nodes connected by 5.2 million weighted (non-binary) ties, resulting in an average degree $\langle k \rangle \approx 3.0$. Each time a user initiated or received a call or a text message, the location of the tower routing the communication was recorded\cite{gonzalez}. We exploited these records to assign each individual to the location where they conducted most of their cell phone communication, which for most individuals is likely to correspond to the location of their home or work. This resulted in one coordinate pair $(x_i, y_i)$ per user, which enabled us to define the geographic distance for any user pair as $d_{ij} = d_{ji} = \sqrt{(x_i - x_j)^2 + (y_i - y_j)^2}$. We used this to compute the probability of a call-tie and the probability of a text-tie as a function of distance (Fig.~2). 

\begin{figure}[!ht]
\begin{center}
\includegraphics[width=0.55\linewidth]{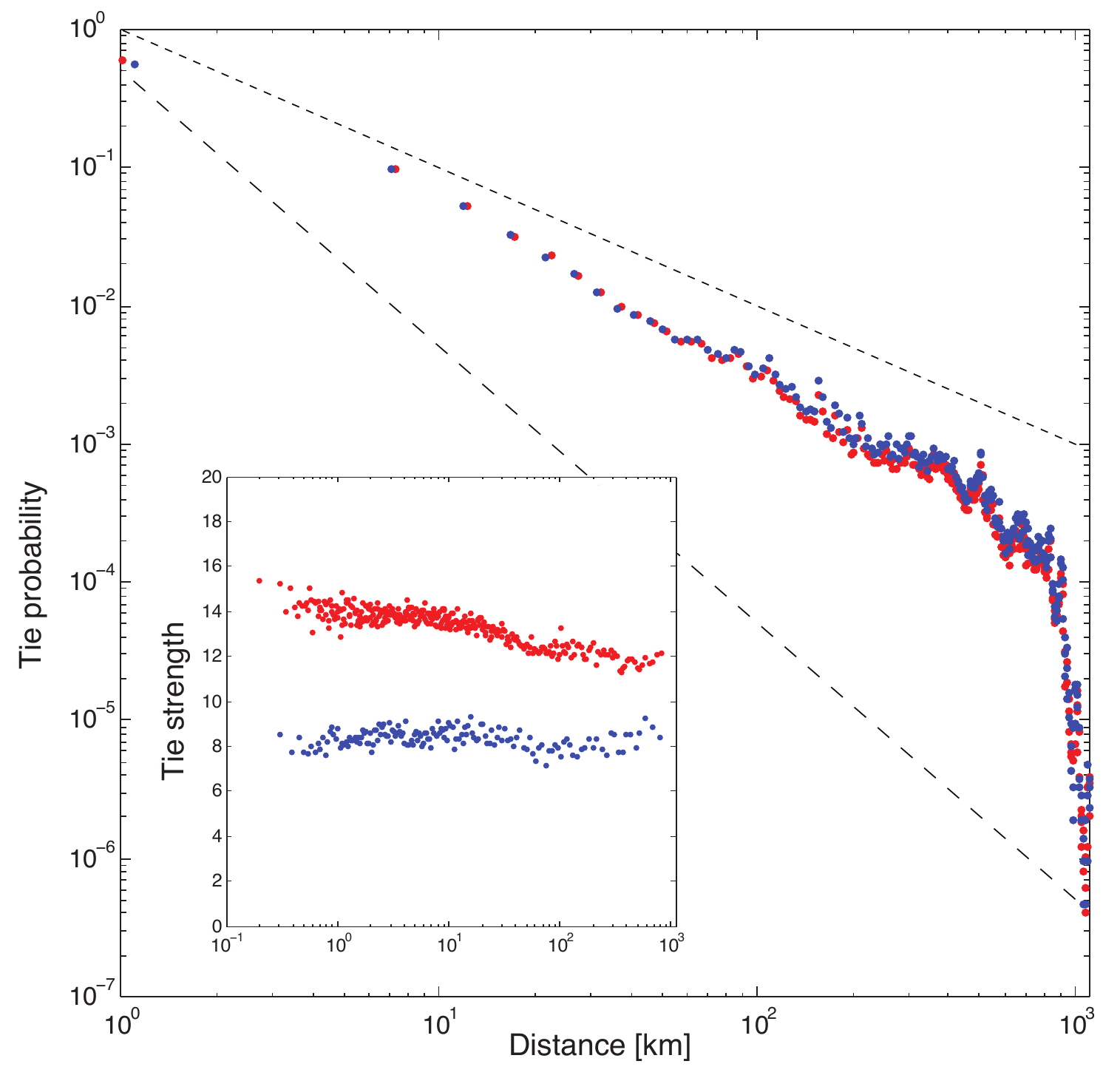}
\end{center}
\caption{{\bf The probability of having a tie decreases as a function of distance.} Two limiting cases, corresponding to exponents one and two, are shown as dashed lines. Note that if geography played no role, we would expect $P(d)$ to be independent of distance $d$, resulting in a horizontal line in this plot. Inset: Tie strength, in contrast to the communication probability, is nearly flat with distance, although there is a minor decreasing trend for voice-ties. }
\end{figure}

Although from the point of view of technology there is very little difference between placing a short-distance or long-distance communication (for either voice or text), we find that the probability of communication is strongly related to the distance between the individuals, and it decreases by approximately five orders of magnitude as distance increases from 1~km to 1,000~km. The behavior of voice-ties and text-ties is essentially identical. The average distance between two connected nodes is 42~km for voice ties and 51~km for text ties. The decay of the tie probability approximately follows a power-law of the form $P(d) \sim d^{- \alpha}$, before it falls due to reaching the physical boundaries of the system. We used the maximum-likelihood method\cite{clauset} to estimate both the exponent $\alpha$ and the lower bound $d_{\min}$ from which the power-law holds, and obtained $\alpha \approx 1.58$ for voice ties and $\alpha \approx 1.49$ for text ties, with the lower bounds estimated at 7.1~km and 4.1~km, respectively. In estimating these parameters, we constrained our search to ties whose distance was less than 800~km to avoid boundary effects, still leaving us over 99\% of the ties.

We define tie strength $w_{ij} = w_{ji}$ as the number of interactions between nodes $i$ and $j$, and it can quantified either as the number of calls between the two nodes or, alternatively, as the number of text messages between them. Interestingly, while geography is so strongly associated with the existence of a tie, tie strength varies only weakly with distance and is similar for both text and voice (Fig.~2 inset).

\subsection*{Community Interaction Structure and Geography}
It is clear that ties or dyads should be the building blocks of social groups or communities, but what constitutes a community and how it should be identified needs to be specified. We detect topological communities using the method of modularity maximization, which measures how well a given partition of a network compartmentalizes its communities\cite{commreview,fortunato,newman1, newman2} (see Methods for details). For this purpose, we combine voice-ties and text-ties into one network. 

Next, we examined how the topological centrality of nodes within communities is associated with their physical centrality. Given the community membership of each individual, we computed the geographical center $(X_s, Y_s)$ of community $s$ using $X_s = (1/n_s) \sum_{i \in C_s} x_i$ and $Y_s = (1/n_s) \sum_{i \in C_s} y_i$, where $n_s$ is the number of members (nodes) in the community. We measured topological centrality using betweenness centrality, whereas physical centrality was measured as the distance from a node to the geographic center of its corresponding community. Given that both betweenness centrality and the physical span of communities increase as a function of community size, we normalized these quantities by considering their percentile values, instead of dealing with their absolute values. (Note that while betweenness centrality can be normalized to be independent of network size, there is no similar normalization available for the physical distances.) We included communities whose size varied between 10 and 1,000 nodes. While the  community detection algorithm found communities that were significantly larger than this upper bound, we deemed them to be too large to be taken as social communities. Including communities smaller than 10 led to discretization effects when computing percentiles.

In historically relevant social arrangements, one might expect the two measures of betweenness centrality and geographic distance from the community center to be strongly correlated, but here we found essentially no correlation between them (Fig.~3). Pearson's linear correlation coefficient between these two measures, both taken as percentiles, was -0.07 (we obtained 0.05 if communities smaller than 10 were also included). Therefore, there seems to be no relationship between topological centrality and physical centrality of nodes within communities in this network.

\begin{figure}[!ht]
\begin{center}
\includegraphics[width=0.5\linewidth]{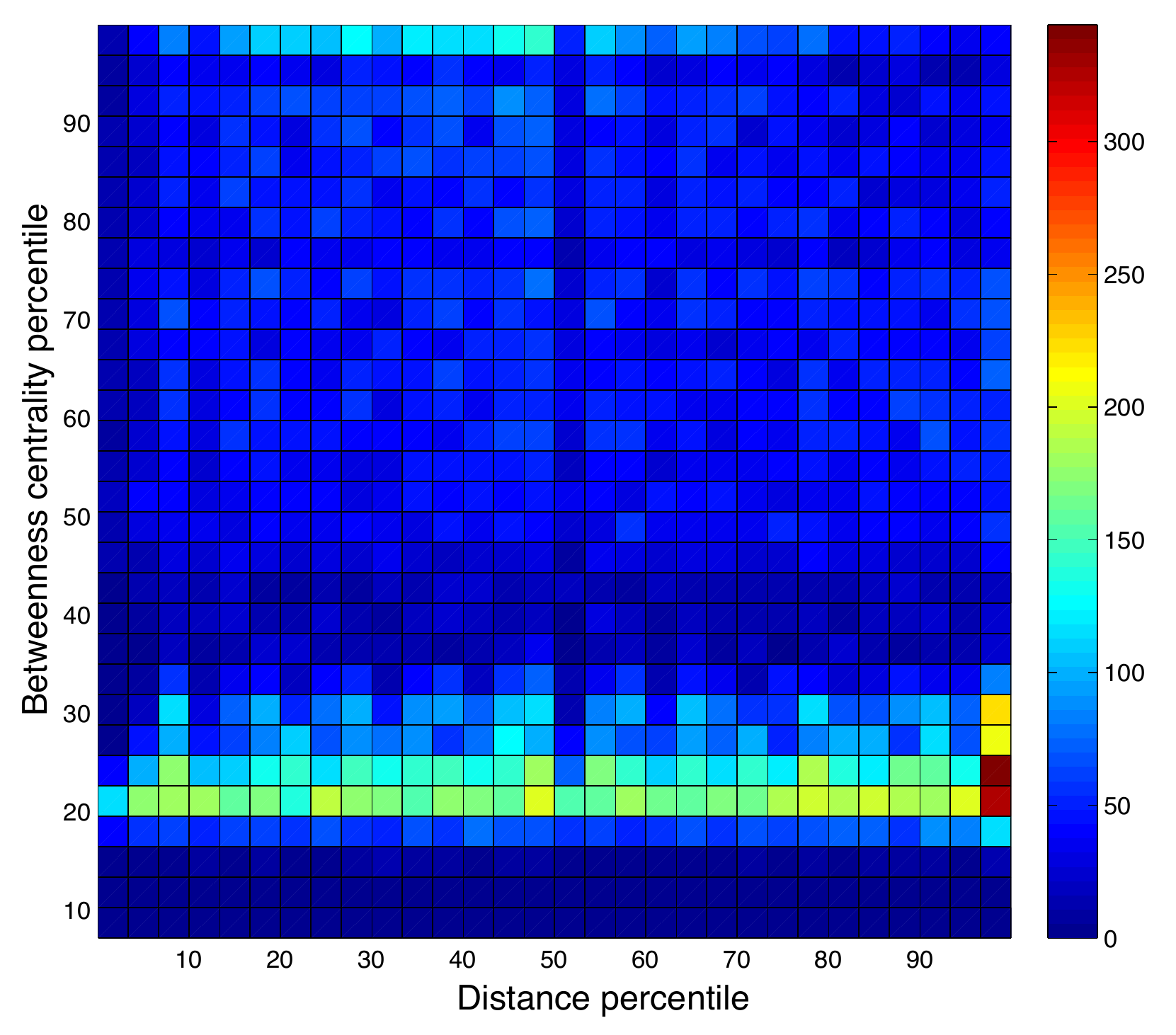}
\end{center}
\caption{{\bf Relationship between topological centrality and geographic centrality.} 42,123 nodes in communities varying in size from 10 to 1,000 are examined, and both quantities are measured in terms of percentiles. The number of observations (nodes) that lie within each bin is indicated by its color.}
\end{figure}

Next, we characterized the overall geographical shape of the communities by defining the \emph{geographical span} for a given community $s$ as
\begin{equation}
D(s) = (1/n_s) \sum_{i \in C_s} \sqrt{(X_s - x_i)^2 + (Y_s - y_i)^2},
\end{equation}
where $D$ is measured in units of distance, and large values of $D$ indicate that the members of the community are geographically spread out. We found an upward trend that persisted with a leveling off until, surprisingly, a large bump occurred for communities in excess of 30 nodes (Fig.~4).

\begin{figure}[!ht]
\begin{center}
\includegraphics[width=0.55\linewidth]{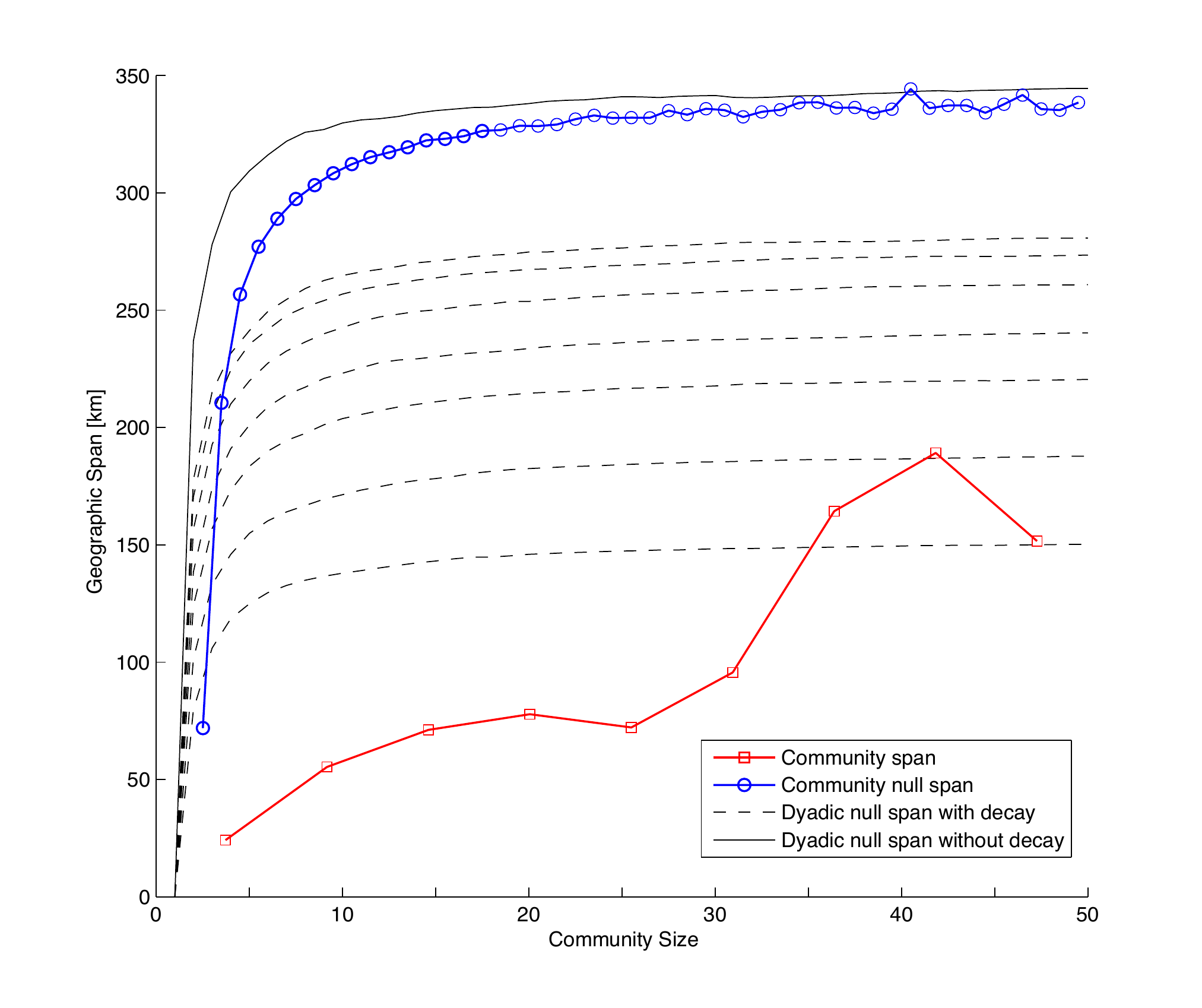}
\end{center}
\caption{{\bf Average observed geographic community span $D$ (red) and average geographic community null span ${D}_{c}$ (blue).} The dyadic null span with decay, denoted by $D_{d}$, incorporates the decay of the connection probability as a function of distance for various values of the scale parameter, shown as dashed lines. The solid black line is the dyadic null span without decay. Both are measured in kilometers. We observe large deviations from both null models, which can be quantified as the areas between the empirical curve and any of the null curves.}
\end{figure}

To put this result in a context, we introduced two null models. In the \emph{community null model}, instead of using the true geographical coordinates $(x_i, y_i)$ of community members, we draw the $(x_i, y_i)$ coordinate pairs uniformly at random from the underlying distribution of all coordinate pairs, keeping a given $x$-coordinate coupled with the associated $y$-coordinate, resulting in the quantity $D_{c}$. If $D_{c}(s) = D(s)$, this would suggest that the members of the community are randomly scattered in the country, i.e., regardless of being members of the same community, they are not geographically proximate. As shown in (Fig.~4), the real community span is much smaller than the span of the null community. What is especially notable is the constraining role of geography for small communities. As community sizes increase, say, from five to ten individuals, the value of the null span increases dramatically from about 70~km to about 300~km, quantifying the expected growth in geographical span if the impact of geography could, somehow, be turned off. Instead, we observe relatively modest growth for the empirical span $D$, which for communities of size ten reaches a value of just 50~km, and stays relatively unchanged until communities exceed 30 in size.

The community null model does not incorporate our earlier finding that the connection probability decays with distance as $P(d) \sim d^{-1.5}$. We next asked whether  this decay, coupled with the concentration of populations in cities, might explain the observed bump. To account for this possibility, we introduce the \emph{dyadic null model}. The algorithm starts by picking one location, uniformly at random, as the geographical center of the community. It then samples other locations, again uniformly at random, and computes the probability for there to be a tie between the center of the community and the current location, where the probability distribution is assumed to follow a power-law with exponent $\alpha = 1.5$. To determine whether the current location is included in the community, the algorithm performs a Bernoulli trial with the given probability, and this continues until we have 50 members in the community. We compared the result of the dyadic null model without decay, achieved by accepting each trial location for inclusion, to the community null span. Apart from a slight horizontal shift, the two null models produce very similar outcomes. We then consider the dyadic null model with decay, varying the value of the scale parameter $d_{\min}$, running each simulation 1,000 times. Although the numerical values are not comparable between the dyadic null model and the community null model, the former demonstrates that inclusion of the decay of the connection probability with distance yields a smooth curve for the span. In particular, decay with distance cannot explain the observed bump.

We also explored the spatial distribution of the nodes within a community. In general, the nodes of a given community need not be distributed spatially uniformly. To quantify this ``clumpiness'' of a community, we wanted to determine the number of spatial clusters making up the community. We used $k$-means clustering\cite{kmeans} which aims to partition the set of data points into $k$ clusters such that each point belongs to the cluster with the nearest mean. Since the number of clusters $k$ is given as input to the method, it can be seen as a model parameter, and it needs to be determined separately. At the extreme ends, one could assign every point to a single cluster, an approach likely to result in a large error measure, or one could assign every point to its own cluster, leading to zero error. We used the Akaike Information Criterion to determine the optimal value for $k$\cite{akaike}.

\begin{figure}[!ht]
\begin{center}
\includegraphics[width=0.55\linewidth]{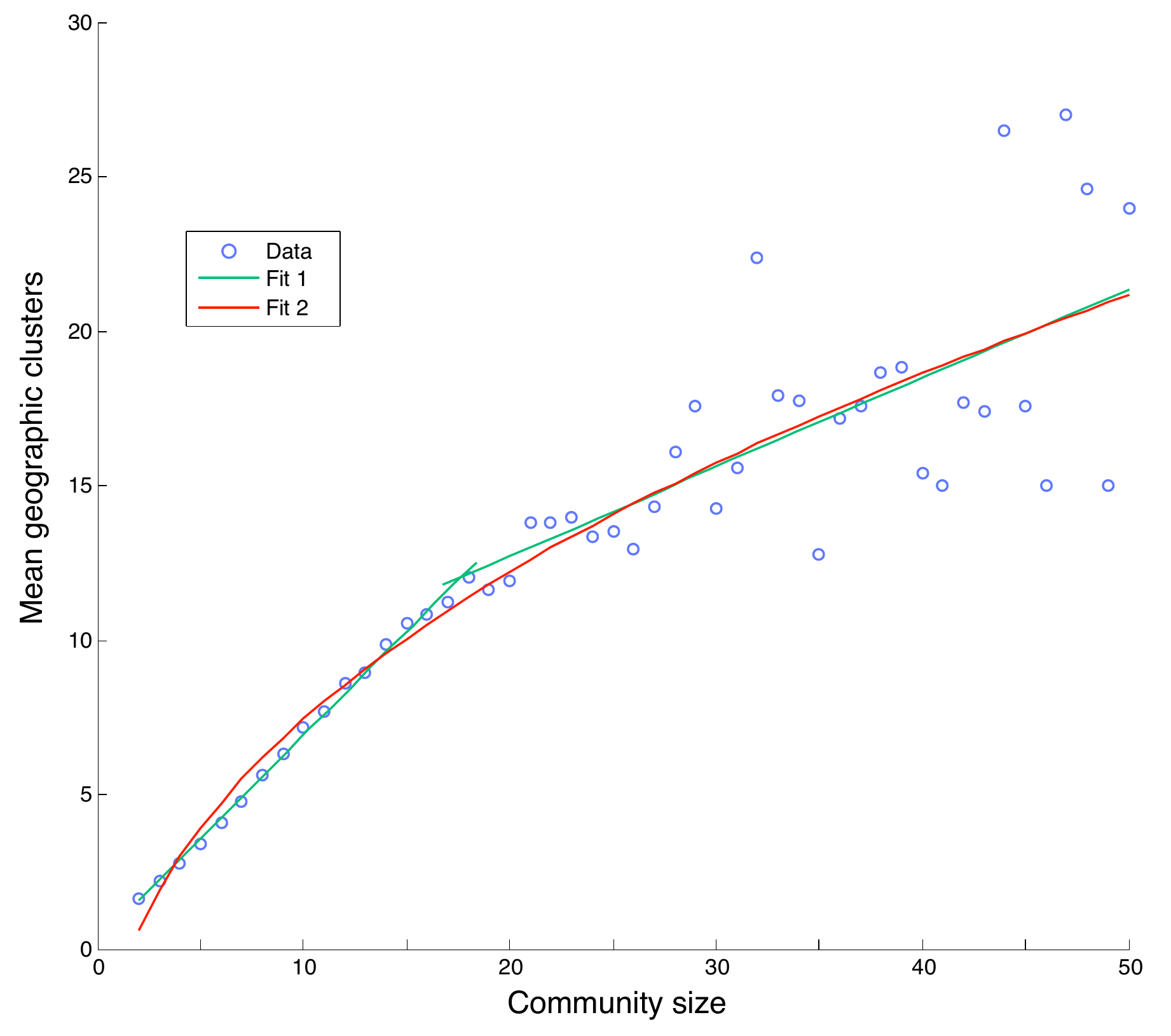}
\end{center}
\caption{{\bf The average number of spatial clusters for empirical data,  versus topological (network) community size.} Clusters are detected using the $k$-means algorithm with the Akaike Information Criterion. We fit two models to data. First, a linear model $y = a_1 + a_2 x$ was fit in two parts, shown in green, as well as a non-linear model $y = b_1 + b_2 x^{b_3}$, shown in red. We obtained the values $a_2 = 0.67$ for the first slope and $a_2 = 0.29$ for the second slope of the linear fits, and $b_3 = 0.43$ for the exponent of the non-linear model, implying approximately square-root behavior.}
\end{figure}

We found that the number of spatial clusters increases linearly with community size, until communities of about size 20, when the behavior appears to change (Fig.~5). The increase in community span for communities larger than 20, without a comparable increase in the number of spatial clusters, suggests a threshold in structure and behavior based on community size. Based on linear fits to data, the addition of an extra community member causes, on average, a marginal increase of 0.67 spatial clusters in small (few members) communities, whereas in large (many members) communities, the addition of an extra community member causes, on average, a marginal increase of 0.29 spatial clusters. For example, an increase from 5 to 15 members increases the number of spatial clusters by 6.1, whereas an increase from 25 to 35 members results in an increase of 2.7 clusters. Communities therefore seem to grow initially by recruiting spatially more distant clusters, but less and less so as the communities get bigger.

\section*{Discussion}
Our findings on the geographic decay of ties differ from those obtained for a network constructed from the customers of a Belgian mobile operator. Using zip codes provided for billing purposes to compute distances between individuals, Lambiotte {\it et al.} showed that the probability for two individuals to be connected decays as $P(d) \sim d^{-2}$, which led them to suggest that the decay follows a so-called gravity model\cite{lambiotte}. Our result, essentially showing that $P(d) \sim d^{-1.5}$, differs for various possible reasons: we used the maximum-likelihood technique to estimate the value of the exponent\cite{clauset}; our range of distances is larger ($800$ km vs. $100$ km), allowing for more statistical power; we used the location of maximal phone use as opposed to the location of the billing address (which is often not reliable); and the population density in our target country is significantly lower than that of Belgium. Our result that tie strength does not vary with distance is complementary to the finding of Lambiotte {\it et al.}, who report that the average duration of phone calls increases with distance, reaching a plateau around 40~km. Therefore, while the number of calls made to long-distance individuals friends is slightly smaller than those made to short distance friends, the average duration may be twice as long\cite{lambiotte}.

Communities appear to have particular properties in relationship to geography, properties that are distinct from the underlying interactions between pairs of individuals. Geography constrains group formation in important ways that nevertheless differ from the way it constrains dyadic interactions. On the one hand, comparison of topological and geographical centrality of nodes within communities demonstrated that the two are essentially uncorrelated. On the other hand, we find that the geographic shapes of social groups, measured in terms of geographic span and spatial clustering, vary in regular ways with the size of the group. For small communities, as their size increases, their expected geographic span increases smoothly at first, but then experiences a sudden bounce as the community size reaches about 30 members. To exemplify this behavior, an increase in community size from 10 to 20 members is associated with an increase in span by about 40\%, whereas, in contrast, an increase from 30 to 40 members leads to an increase of about 100\% in geographic span. This suggests that the tendency of human groups to remain geographically cohesive gradually gives in as the group size exceeds 30. Similarly, the number of clusters within a single topological group also increases with community size. Intriguingly, the number 30 is also close to the optimal group size for which cooperation in social dilemma situations, modeled, for example, by the public goods game, is maximized \cite{perc,skyrms}.

Just as the structures of observed social interactions may be compared to randomized networks, the observed localities of individuals in communities may be compared to randomized locations. Indeed, if social ties could be formed without consideration for the underlying geography, we would expect the tie probability to be independent of distance, and the geographic span of groups to follow the proposed null models closely. However, we observe neither of the two. This demonstrates that network ties and network communities, in this context, do not behave as if they were in well mixed populations, suggesting that geography continues to maintain its power as a compartmentalizing factor. Thus, the assumption of perfect mixing of individuals, sometimes made in the study of infectious disease or technology diffusion in humans, does not then appear to hold either at the topological or at the geographic level.

The extent to which a spreading process follows the assumptions of well-mixed populations often depends on a number of conditions, including the nature of the spreading process. For example, network models can better account for the spread of diseases that spread via the formation of a physical tie (such as STD's) than those that spread by simple proximity (like the common cold). This is illustrated in a mobile phone context by Wang {\it et al.}\cite{wang}, who find that the nature of the spreading process, and its dependence on proximity, clearly affect the dynamics of the spreading. Similarly, localization of interacting proteins within the geography of the cell can explain certain disease associations\cite{park}. In addition, other work has suggested that the diffusion-like movement of people alone can often explain how a pathogen spreads, such as the plague in medieval Europe\cite{noble}. On the other hand, with the onset of air travel, pathogens are not constrained in the same way, as the epidemics of SARS and H1N1 documented\cite{hufnagel,vespignani}.

Ideally, models of the flow of pathogens or information through human populations would account for the simultaneous roles of geographic and network constraints, and our work helps shed light on the intersecting relationship between the two. Future work will explore the complex interrelationship between network topology and geography and their joint importance in understanding how phenomena spread through populations.

\section*{Methods}
All networks were constructed from four weeks of anonymized mobile phone call and text messaging data from an operator based in an unnamed European country. Only interactions that took place between customers of the operator were considered, and only individuals who made at least two calls were included as nodes. To filter out sporadic calls and texts that are unlikely to correspond to meaningful social interactions, we required there to be a minimal level of reciprocation for a tie to be included in the network; each person had to initiate at least one interaction, where the initial transaction could be either a call or a text, and this could be reciprocated by either a call or a text.

We detected topological communities using the popular method of modularity maximization\cite{commreview,fortunato,newman1, newman2} in the following manner. We first converted the original network consisting of directed voice calls and text messages into a symmetric unweighted network, effectively combining the two modes of interaction. We then proceeded to maximize modularity defined as 
\begin{equation}
Q = \frac{1}{2w}\sum_{i,j} \left[ A_{ij} - \frac{k_i k_j}{2w} \right] \delta(c_i, c_j), 
\end{equation}
where the adjacency matrix element $A_{ij}$ denotes the presence ($A_{ij} = 1$) or absence ($A_{ij} = 0$) of a connection between nodes $i$ and $j$, $k_i$ is the degree of node $i$, $w$ the total weight of the edges in the network, $c_i$ the community assignment of node $i$, and $\delta(c_i, c_j)$ is the Kronecker delta function, which is unity if and only if $c_i = c_j$, otherwise it is zero. Modularity measures the difference between the total fraction of edges that fall within groups versus the fraction one would expect by chance. A common null model, sometimes called the Newman-Girvan null model, is codified by the $k_ik_j/(2w)$ term, and it takes degree heterogeneity into account by preserving the expected degree distribution. High values of $Q$  indicate network partitions in which more of the edges fall within groups than expected by chance. While maximizing modularity is known to be an NP-hard problem\cite{np}, there are numerous computational heuristics available\cite{commreview,fortunato}. Since we are dealing with networks consisting of millions of nodes, we chose the Louvain method for its computational efficiency\cite{blondel}. 

\section*{Acknowledgments}
We acknowledge A.~Zaslavsky for useful discussions about the Akaike Information Criterion. JPO, SA, and NAC are supported by the National Institute on Aging (grant P01 AG-031093); ALB by the Office of Naval Research (grant ONR N000141010968), the Network Science Collaborative Technology Alliance (grant ARL NS-CTA W911NF-09-2-0053),  the Defense Threat Reduction Agency (grants DTRA BRBAA08-Per4-C-2-0033 and DTRA WMD BRBAA07-J-2-0035), the National Science Foundation (grant NSF BCS-0826958), and the James S.~McDonnell Foundation (grant JSMF 220020084).


\begin{thebibliography}{10}
\providecommand{\url}[1]{\texttt{#1}}
\providecommand{\urlprefix}{URL }
\expandafter\ifx\csname urlstyle\endcsname\relax
  \providecommand{\doi}[1]{doi:\discretionary{}{}{}#1}\else
  \providecommand{\doi}{doi:\discretionary{}{}{}\begingroup
  \urlstyle{rm}\Url}\fi
\providecommand{\bibAnnoteFile}[1]{%
  \IfFileExists{#1}{\begin{quotation}\noindent\textsc{Key:} #1\\
  \textsc{Annotation:}\ \input{#1}\end{quotation}}{}}
\providecommand{\bibAnnote}[2]{%
  \begin{quotation}\noindent\textsc{Key:} #1\\
  \textsc{Annotation:}\ #2\end{quotation}}
\providecommand{\eprint}[2][]{\url{#2}}

\bibitem{newman}
Newman MEJ (2008) The physics of networks.
\newblock Physics Today 61: 33-38.


\bibitem{caldarelli}
Caldarelli G (2007) Scale-Free Networks: Complex Webs in Nature and Technology.
\newblock Oxford University Press, 1st edition.


\bibitem{mendes}
Dorogovtsev SN, Mendes JFF (2003) Evolution of Networks: From Biological Nets
  to the Internet and WWW.
\newblock Oxford University Press, 1st edition.


\bibitem{albert}
Albert R, Barab\'asi AL (2002) Statistical mechanics of complex networks.
\newblock Rev Mod Phys 74: 47-97.


\bibitem{structure}
Newman MEJ, Barab\'asi AL, Watts DJ (2006) The Structure and Dynamics of
  Networks.
\newblock Princeton University Press, 1st edition.


\bibitem{commreview}
Porter MA, Onnela J-P, Mucha PJ (2009) Communities in networks.
\newblock Notices of the American Mathematical Society 56: 1082.


\bibitem{fortunato}
Fortunato S (2010) Community detection in graphs.
\newblock Physics Reports 486: 75-174.


\bibitem{freemanbook}
Freeman LC (2004) The Development of Social Network Analysis: A Study in the
  Sociology of Science.
\newblock Vancouver, Canada: Empirical Press.


\bibitem{onnela}
Onnela J-P et~al. (2007) tructure and tie strengths in mobile communication networks.
\newblock P Natl Acad Sci USAS 104: 7332-7336.


\bibitem{gonzalez}
Gonz\'alez MC, A HC, Barab\'asi AL (2008) Understanding individual human
  mobility patterns.
\newblock Nature 453: 779-782.


\bibitem{libennowell}
Liben-Nowell D, et~al. (2005) Geographic routing in social networks.
\newblock P Natl Acad Sci USA 102: 11623-11628.


\bibitem{lambiotte}
Lambiotte R, et~al. (2008) Geographical dispersal of mobile communication networks.
\newblock Physica A 387: 5317-5325.


\bibitem{palla}
Palla G, Barab\'asi AL, Vicsek T (2007) Quantifying social group evolution.
\newblock Nature 446: 664-667.


\bibitem{ahn}
Ahn YY, Bagrow JP, Lehmann S (2010) Link communities reveal multiscale complexity in
  networks.
\newblock Nature 466: 761-764.


\bibitem{clauset}
Clauset A, Shalizi CR, Newman MEJ (2009) Power-law distributions in empirical
  data.
\newblock SIAM Review 51: 661-703.


\bibitem{newman1}
Newman MEJ (2003) Mixing patterns in networks.
\newblock Phys Rev E 67: 026126.


\bibitem{newman2}
Girvan M, J Newman MEJ (2002) Community structure in social and biological networks.
\newblock P Natl Acad Sci USA 99: 7821.


\bibitem{kmeans}
Gan G, Ma C, Wu J (2007) Data Clustering: Theory, Algorithms, and Applications.
\newblock SIAM.


\bibitem{akaike}
Akaike H (1974) A new look at the statistical model identification.
\newblock IEEE Transactions on Automatic Control 19: 716-723.


\bibitem{perc}
Szolnoki A, Perc M (2010) Impact of critical mass on the evolution of cooperation in spatial public goods games. 
\newblock Phys Rev E 81: 057101.

\bibitem{skyrms}
Skyrms B (2004) The stag hunt and the evolution of social structure.
\newblock Cambridge University Press.

\bibitem{wang}
Wang P, Gonz\'alez MC, Hidalgo CA, Barab\'asi AL (2009) Understanding the
  spreading patterns of mobile phone viruses.
\newblock Science 324: 1071-1076.

\bibitem{park}
Park S The importance of protein subcellular localization for disease
  profiling.


\bibitem{noble}
Noble JV (1974) Geographic and temporal development of plagues.
\newblock Nature 250: 726-729.


\bibitem{hufnagel}
Hufnagel L, Brockmann D, Geisel T (2004) Forecast and control of epidemics in a
  globalized world.
\newblock P Natl Acad Sci USA 101.


\bibitem{vespignani}
Colizza V, Barrat A, Barthelemy M, Valleron AJ, Vespignani A (2007) Modeling
  the worldwide spread of pandemic influenza: baseline case and containment
  interventions.
\newblock PLoS Med 4: e13.


\bibitem{np}
Brandes U, et~al. (2008) On modularity clustering.
\newblock IEEE Transactions on Knowledge and Data Engineering 20: 172-188.


\bibitem{blondel}
Blondel VD, Guillaume JL, Lambiotte R, Lefebvre E (2008) Fast unfolding of communities in large network.
\newblock Journal of Statistical Mechanics: Theory and Experiment : P10008.

\end{thebibliography}
\end{document}